\documentclass[12pt]{article}

\newcommand{\m}{\mathrm}
\newcommand{\be}{\begin{equation}}
\newcommand{\ee}{\end{equation}}
\newcommand{\ba}{\begin{eqnarray}}
\newcommand{\ea}{\end{eqnarray}}

\usepackage{graphicx}
\usepackage{amssymb}
\usepackage{amsmath}
\usepackage[T1]{fontenc} 
\usepackage[ansinew]{inputenc} 
\usepackage[nosort]{cite}
\newcommand{\inbar}{\vrule height1.57ex width.4pt depth0pt}
\newcommand{\SW}{\relax{\hbox{$\ \inbar\kern-.285em{\rm S}$}}}

\begin{document}
\thispagestyle{empty}
\begin{center}

\null \vskip-1truecm \vskip2truecm

{\Large{\bf \textsf{Fragmentation of AdS$_5$-Kerr Black Holes}}}

{\large{\bf \textsf{}}}

{\large{\bf \textsf{}}}

\vskip1truecm

{\large \textsf{Brett McInnes}}

\vskip1truecm

\textsf{\\  National
  University of Singapore}

\textsf{email: matmcinn@nus.edu.sg}\\

\end{center}
\vskip1truecm \centerline{\textsf{ABSTRACT}} \baselineskip=15pt
\medskip

Black hole spacetimes asymptotic to five-dimensional anti-de Sitter spacetime are of great interest in connection with the string-gauge duality. In the rotating case, such black holes tend to become unstable, in several different ways, if their specific angular momenta fall in certain ranges. Here we consider the well-known Emparan-Myers fragmentation instability for singly rotating AdS$_5$-Kerr black holes, paying particular attention to the case where the specific angular momentum exceeds the asymptotic AdS$_5$ curvature length scale.

\newpage

\addtocounter{section}{1}
\section* {\large{\textsf{1. AdS$_5$-Kerr Black Holes}}}
In the string-gauge duality \cite{kn:casa,kn:nat,kn:bag}, five-dimensional asymptotically AdS black holes are dual to four-dimensional, thermal systems, and hence play a central role. Perhaps the most interesting examples are the AdS$_5$-Kerr black holes\footnote{For simplicity, we focus here on five-dimensional black holes rotating about a single axis. Henceforth, we take ``AdS$_5$-Kerr'' to mean this case. Ideally, one would study AdS$_5$-Kerr-Newman black holes, but that is not yet possible at the level of detail we want here. See \cite{kn:emparan} for this (difficult and poorly understood) extension.}, and this is the case on which we will focus here.

It is possible for an AdS$_5$-Kerr black hole to satisfy classical (weak) Censorship (which one expects to hold in the AdS case \cite{kn:weak,kn:horsant,kn:suvrat,kn:bala,kn:sean}, whatever its status may be in the asymptotically flat or de Sitter contexts) and yet have an arbitrarily large specific angular momentum, even if the mass is fixed \cite{kn:96,kn:98}. To be precise, Censorship for these black holes excludes (for a given mass) only a finite \emph{band} of values (around unity\footnote{The limiting case, $\mathcal{A}/L \rightarrow 1$, has been discussed in \cite{kn:klem,kn:hen,kn:supe}; we will not consider it here.}) for $\mathcal{A}/L$: see below for the details.

When one embeds these objects in string theory, however, the picture changes in a remarkable and suggestive manner. One now finds \cite{kn:96} that the specific angular momentum $\mathcal{A}$ of an AdS$_5$-Kerr black hole must satisfy, if $L$ is the asymptotic AdS$_5$ curvature scale,
\begin{equation}\label{B}
\mathcal{A}\;\leq \;2\,\sqrt{2}\,L,
\end{equation}
since otherwise the system is subject, far away from the event horizon, to a non-perturbative BPS brane pair-production instability of the kind studied by Seiberg and Witten \cite{kn:seiberg}.

We regard (\ref{B}) as the fundamental bound on the specific angular momenta of uncharged AdS$_5$ black holes in string theory. Consequently, we take the view that all black holes which \emph{do} satisfy (\ref{B}) must be considered as being potentially of physical interest. In view of our comments above, we will concentrate on uncharged AdS$_5$ black holes that satisfy \emph{both} Censorship and (\ref{B}).

According to our discussion, these fall (in general) into two classes, each characterised by a finite range of values for $\mathcal{A}/L$. One of these ranges is from zero up to some value strictly smaller than unity --$\,$ let us call the corresponding black holes \emph{cisunital} black holes --$\,$ while the other is from some value strictly greater than unity, up to $2\sqrt{2}$; we will describe the corresponding black holes as \emph{transunital} black holes.

Rotating black holes are endangered by two important instabilities, quite apart from the Seiberg-Witten instability. These are the effects associated with \emph{superradiance} \cite{kn:super} and the distortions of the shape of the event horizon arising at high angular momenta, which can lead to black hole \emph{fragmentation} \cite{kn:empmy,kn:pau}.

We found in \cite{kn:98} that some cisunital black holes do suffer from a superradiant instability, and so do most transunital black holes: most, \emph{but not all}. In this work, we will ask whether the survivors can also survive the Emparan-Myers fragmentation instability. Again, we will see that the evidence suggests that some cisunital, and most transunital, black holes are unstable in this manner, \emph{but not all}. Requiring stability against all these effects strongly restricts the possible specific angular momenta of AdS$_5$-Kerr black holes. According to the string-gauge duality, it therefore imposes strong restrictions on the specific angular momenta possible in the dual four-dimensional system.

We begin with a review of the general properties of uncharged asymptotically AdS$_5$ black holes.

\addtocounter{section}{1}
\section* {\large{\textsf{2. Basic Properties of Cis/Transunital AdS$_5$-Kerr Black Holes}}}
The AdS$_5$-Kerr metric \cite{kn:hawk,kn:cognola,kn:gibperry}, for a singly rotating, uncharged black hole, is given by
\begin{flalign}\label{C}
g\left(\m{AdS_5K}\right)\; = \; &- {\Delta_r \over \rho^2}\left[\,\m{d}t \; - \; {a \over \Xi}\,\m{sin}^2\theta \,\m{d}\phi\right]^2\;+\;{\rho^2 \over \Delta_r}\m{d}r^2\;+\;{\rho^2 \over \Delta_{\theta}}\m{d}\theta^2 \\ \notag \,\,\,\,&+\;{\m{sin}^2\theta \,\Delta_{\theta} \over \rho^2}\left[a\,\m{d}t \; - \;{r^2\,+\,a^2 \over \Xi}\,\m{d}\phi\right]^2 \;+\;r^2\cos^2\theta \,\m{d}\psi^2 ,
\end{flalign}
where
\begin{eqnarray}\label{D}
\rho^2& = & r^2\;+\;a^2\cos^2\theta, \nonumber\\
\Delta_r & = & \left(r^2+a^2\right)\left(1 + {r^2\over L^2}\right) - 2M,\nonumber\\
\Delta_{\theta}& = & 1 - {a^2\over L^2} \, \cos^2\theta, \nonumber\\
\Xi & = & 1 - {a^2\over L^2}.
\end{eqnarray}
Here $L$ is the background AdS curvature length scale as above, $t$ and $r$ are as usual, $\phi$ and $\psi$ run from $0$ to $2\pi$, and $\theta$ runs from $0$ to $\pi/2$.

The quantities $a$ and $M$, together with $L$, describe the geometry of the black hole spacetime. They are \emph{not} equal in general to the specific angular momentum and physical mass of the black hole \cite{kn:gibperry}.

If $\mathcal{M}$ is the physical mass, then
\begin{equation}\label{E}
\mathcal{M}\;=\;{\pi M \left(2 + \Xi\right)\over 4\,\ell_{\textsf{P}}^3\,\Xi^2},
\end{equation}
where $\ell_{\textsf{P}}$ is the bulk Planck length. It is often more convenient to use a dimensionless version of this quantity, defined by
\begin{equation}\label{L}
\mu \equiv {8\ell_{\textsf{P}}^3\mathcal{M}\over \pi L^2}.
\end{equation}
This dimensionless mass is related to the geometric parameter $M$ by
\begin{equation}\label{LLL}
\mu \;=\;{2M\over L^2}\,\left({2 + \Xi \over \Xi^2}\right).
\end{equation}
The physical angular momentum of the black hole is given by
\begin{equation}\label{F}
\mathcal{J}\;=\;{\pi M a\over 2\,\ell_{\textsf{P}}^3\,\Xi^2}.
\end{equation}
The angular momentum to (physical) mass ratio $\mathcal{A}$ is therefore given by
\begin{equation}\label{G}
\mathcal{A}\;=\;{2 a \over 2 + \Xi}\;=\;{2 a \over 3 - \left(a^2/L^2\right)}.
\end{equation}
From this one sees that $\mathcal{A}/L$ is a simple monotonically increasing function of $a/L$; $\mathcal{A}/L$ satisfies $\mathcal{A}/L > 1$ if and only if $a/L > 1$. When $\mathcal{A}/L < 1$, it is a little smaller than $a/L$; when $\mathcal{A}/L > 1$, it is somewhat larger (up to twice as large on the physical domain). The fundamental restriction (\ref{B}) above translates to
\begin{equation}\label{H}
a\;\leq\;\sqrt{2}\,L.
\end{equation}

Every uncharged asymptotically  AdS$_5$ black hole can be characterised by either the geometric parameters $(M/L^2, a/L)$, or by the physical characteristics $(\mu, \mathcal{A}/L)$. Each pair can be regarded as coordinates in the first quadrant of the plane; each pair is mapped to the other by a continuous invertible mapping defined by the relations (\ref{LLL}) and (\ref{G}). Often it is convenient to use $(M/L^2, a/L)$, but of course at any point one can transfer to $(\mu, \mathcal{A}/L)$ as needed.

Censorship for these black holes, when combined with (\ref{B}), takes the following form \cite{kn:98}.

One finds that $\mathcal{A}/L$ must lie in one of two possible ranges: either
\begin{equation}\label{I}
{\mathcal{A}\over L} \;<\; \Gamma_{\mu}^- \;<\;1,
\end{equation}
\emph{or}, taking (\ref{B}) into account,
\begin{equation}\label{J}
1\;<\; \Gamma_{\mu}^+ \;<\; \mathcal{A}/L \;<\;2\,\sqrt{2},
\end{equation}
where $\Gamma_{\mu}^+$ and $\Gamma_{\mu}^-$ are given by
\begin{equation}\label{K}
\Gamma_{\mu}^{\pm}\;=\;2\,\sqrt{2}\,\sqrt{\mu + 1}\,{\sqrt{3 + 2\mu \pm \sqrt{9 + 8\mu}}\over 3 + 4\mu \mp \sqrt{9 + 8\mu}};
\end{equation}
The graph of $\Gamma_{\mu}^-$ is shown in Figure 1.
\begin{figure}[!h]
\centering
\includegraphics[width=0.7\textwidth]{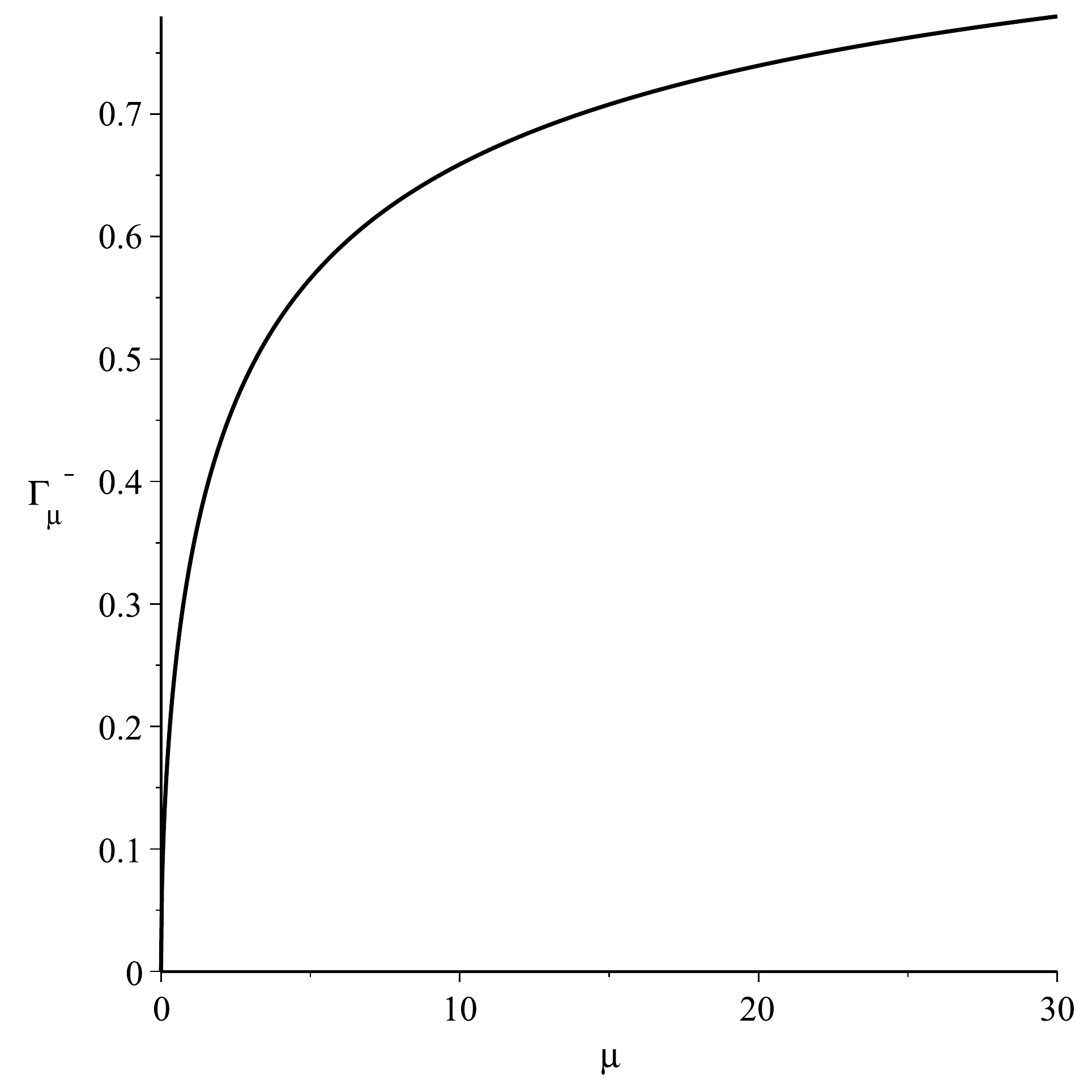}
\caption{Graph of $\Gamma_{\mu}^-$.}
\end{figure}

The bound in (\ref{I}) is continuously related to the asymptotically flat case: (\ref{I}) can be expressed in terms of a power series:
\begin{equation}\label{LANGNAMVEREIN}
\mathcal{A}\;<\;{4\sqrt{2}\over3\sqrt{3\pi}}\,\ell^{3/2}_{\textsf{P}}\mathcal{M}^{1/2}\;-\;{16\sqrt{2}\over 9\sqrt{3\pi^3}}\,{\ell^{9/2}_{\textsf{P}}\mathcal{M}^{3/2}\over L^2}\;+\;\cdots
\end{equation}
The first term on the right corresponds to Censorship for uncharged asymptotically flat ($L \rightarrow \infty$) five-dimensional Kerr (Myers-Perry) black holes; the specific angular momentum is bounded by a multiple of the square root of the mass. There is no analogue of (\ref{J}) in the asymptotically flat case.

When Censorship holds, the event horizon is located at $r = r_{\textsf{H}}$, which can be found by solving $\Delta_r = 0$, that is, by solving the quartic
\begin{equation}\label{LL}
\left(r_{\textsf{H}}^2+a^2\right)\left(1 + {r_{\textsf{H}}^2\over L^2}\right) - 2M\;=\;0.
\end{equation}
Notice that this equation implies that a ``near-extremal'' black hole of this kind has $r_{\textsf{H}} \approx 0$. (It follows \cite{kn:empmy} that there are no exactly extremal black holes of this kind: as extremality is approached, the event horizon eventually disappears, revealing a naked ring singularity.)

One can show that the effect of increasing $\mu$ is to raise $\Gamma_{\mu}^-$ but to lower $\Gamma_{\mu}^+$: the excluded band becomes narrower. Conversely, if $\mu$ is sufficiently small, then $\Gamma_{\mu}^+$ can rise to meet $2\,\sqrt{2}$, so that the range in (\ref{J}) ceases to exist, leaving us only with the more familiar form of Censorship expressed as (\ref{I}). This happens, however, only if $\mu \leq 2$, so generically there are indeed two ranges for $\mathcal{A}/L$ compatible with Censorship.

\emph{Henceforth, we assume that Censorship holds for all asymptotically AdS spacetimes}: see \cite{kn:weak,kn:horsant,kn:suvrat,kn:bala,kn:sean}.

When $\mathcal{A}/L > 1$, we have also $a/L > 1$, so $\Delta_{\theta}$ vanishes at $\theta = \theta_a$, where
\begin{equation}\label{M}
\theta_a\;\equiv \;\arccos\left(L/a\right);
\end{equation}
notice that (\ref{H}) can be expressed as
\begin{equation}\label{MMM}
\theta_a\;\leq \;\pi/4.
\end{equation}
One can readily verify \cite{kn:96} that the ``singularity'' at this value of $\theta$ is just a coordinate ``singularity'', in direct analogy with the event horizon. The analogy continues: just as the signatures of the radial and time coordinates are exchanged when one crosses the event horizon of a Schwarzschild black hole, so also there are signature changes when $\theta = \theta_a$ is crossed, as we will see.

In that connection, we note that whenever $a \neq 0$, these black holes have an \emph{ergosphere}, described by solving the equation
\begin{equation}\label{MM}
-\,\Delta_r\;+\;a^2\sin^2(\theta)\Delta_{\theta}\;=\;0
\end{equation}
for $r$ as a function of $\theta$. Notice that, when $\mathcal{A}/L > 1$, the ergosphere intersects the event horizon along $\theta = \theta_a$, instead of precisely at the poles as happens when $\mathcal{A}/L < 1$ .

When $\mathcal{A}/L < 1$, or for $\theta > \theta_a$ otherwise, the ergosphere lies outside the event horizon, as usual; this means that, on the event horizon in those cases, the coordinate $t$ \emph{has already become spacelike}, similar to $\theta$ and $\phi$. This is the first signature ``flip'' we must take into account.

Let us examine the event horizon more closely. If we restrict the AdS$_5$-Kerr metric to $r = r_{\textsf{H}}$, we obtain
\begin{equation}\label{N}
h_{\textsf{H}}\; = \; {\rho_{\textsf{H}}^2 \over \Delta_{\theta}}\m{d}\theta^2\;+\;{\m{sin}^2\theta \,\Delta_{\theta} \over \rho_{\textsf{H}}^2}\left[a\,\m{d}t \; - \;{r_{\textsf{H}}^2\,+\,a^2 \over \Xi}\,\m{d}\phi\right]^2 \;+\;r_{\textsf{H}}^2\cos^2\theta \,\m{d}\psi^2,
\end{equation}
where $\rho_{\textsf{H}}$ is the value of $\rho$ on the event horizon (so it is a function of $\theta$). Notice that this is a degenerate ``metric'': the rank is (\emph{at most}) 3, not 4. It is positive-definite when $\mathcal{A}/L < 1$, or for $\theta > \theta_a$; this is consistent with the fact that, as explained, $t$ is spacelike on the event horizon in these cases. (None of this has anything to do with the possibility that $\mathcal{A}/L$ might be greater than unity.)

Using $h_{\textsf{H}}$ to compute the circumference $\mathcal{C}$ of the equator ($\theta = \pi/2)$ parametrised by $\phi$, we find
\begin{equation}\label{NN}
\mathcal{C}\;=\;{2\pi\left(r_{\textsf{H}}^2 + a^2\right)\over r_{\textsf{H}}\,|\Xi|}.
\end{equation}
It is important to notice that \emph{either} when the black hole is close to ``extremality'' (which, as explained earlier, means that $r_{\textsf{H}}$ is close to zero) \emph{or} when $\mathcal{A}/L$ is close to unity (which means that $\Xi$ is close to zero), this circumference becomes extremely large\footnote{``Large'' means ``relative to the circumference of a circle on the event horizon oriented in the non-rotating azimuthal direction parametrised by $\psi$'': for example, take the circle of that sort located at $\theta = \pi/4$ (according to (\ref{MMM}), this exists in all cases), which has circumference equal to $\sqrt{2}\pi r_{\textsf{H}}$. Notice that this is \emph{small} near ``extremality''.}. This indicates that the black hole has become flattened along this equatorial plane, and the results of \cite{kn:empmy,kn:pau} for the asymptotically flat case, and of \cite{kn:bog} for the asymptotically AdS case when $\mathcal{A}/L < 1$, suggest that this might in some cases cause an instability. We will show later that this is the case also when $\mathcal{A}/L > 1$.

The circumference is a good indicator of the presence of such behaviour; also, it is the relevant parameter if we need to assess the external ``size'' of the black hole in the plane of rotation, in the sense that any circle in that plane of smaller circumference must be inside the event horizon. Furthermore it is not dependent on the shape of the event horizon in other, irrelevant directions. We therefore use the circumference to define a formal ``radius'' of the black hole simply by
\begin{equation}\label{NORMAL}
R_{\textsf{c}}\;\equiv \;{\mathcal{C}\over 2\pi}\;=\;{\left(r_{\textsf{H}}^2 + a^2\right)\over r_{\textsf{H}}\,|\Xi|};
\end{equation}
clearly $R_{\textsf{c}} \approx r_{\textsf{H}}$ when the angular momentum is small. Of course, no black hole has a ``radius'' in the true geometric sense; we wish to suggest that this \emph{circumferential radius} is the best that can be done in the case of these flattened black holes.

Now let us use $h_{\textsf{H}}$ to determine the entropy of the black hole.

When $0 < \mathcal{A}/L < 1$, or outside $\theta = \theta_a$ when $\mathcal{A}/L > 1$, the area of the event horizon is defined as the area of the three-dimensional surface obtained by fixing the ``time'' coordinate; that is, by setting $t = $ constant. If we do this, then the rank of $h_{\textsf{H}}$ takes its maximal value, 3, so it ceases to be degenerate, and we can use it to compute a non-zero area.  When $\mathcal{A}/L < 1$ this is a straightforward computation, and evaluating the entropy using the Hawking formula we have
\begin{equation}\label{NNN}
S_{\textsf{H}}(\mathcal{A}/L < 1)\; =\; {\pi^2\over 2 \ell_{\textsf{P}}^3}\,{r^2_{\textsf{H}} + a^2\over \Xi}\,r_{\textsf{H}}.
\end{equation}

Using equation (\ref{LL}) to eliminate $r_{\textsf{H}}$, and using (\ref{LLL}) to replace $M$ with the dimensionless mass $\mu$, one can express this as a function of $\mathcal{A}/L$ and $\mu$, or, equivalently and more conveniently, of $a/L$ and $\mu$. One finds that, when the mass $\mu$ is fixed, the entropy is a monotonically decreasing function of $a/L$: see for example Figure 2, which shows the graph for $\mu = 30$.
\begin{figure}[!h]
\centering
\includegraphics[width=0.7\textwidth]{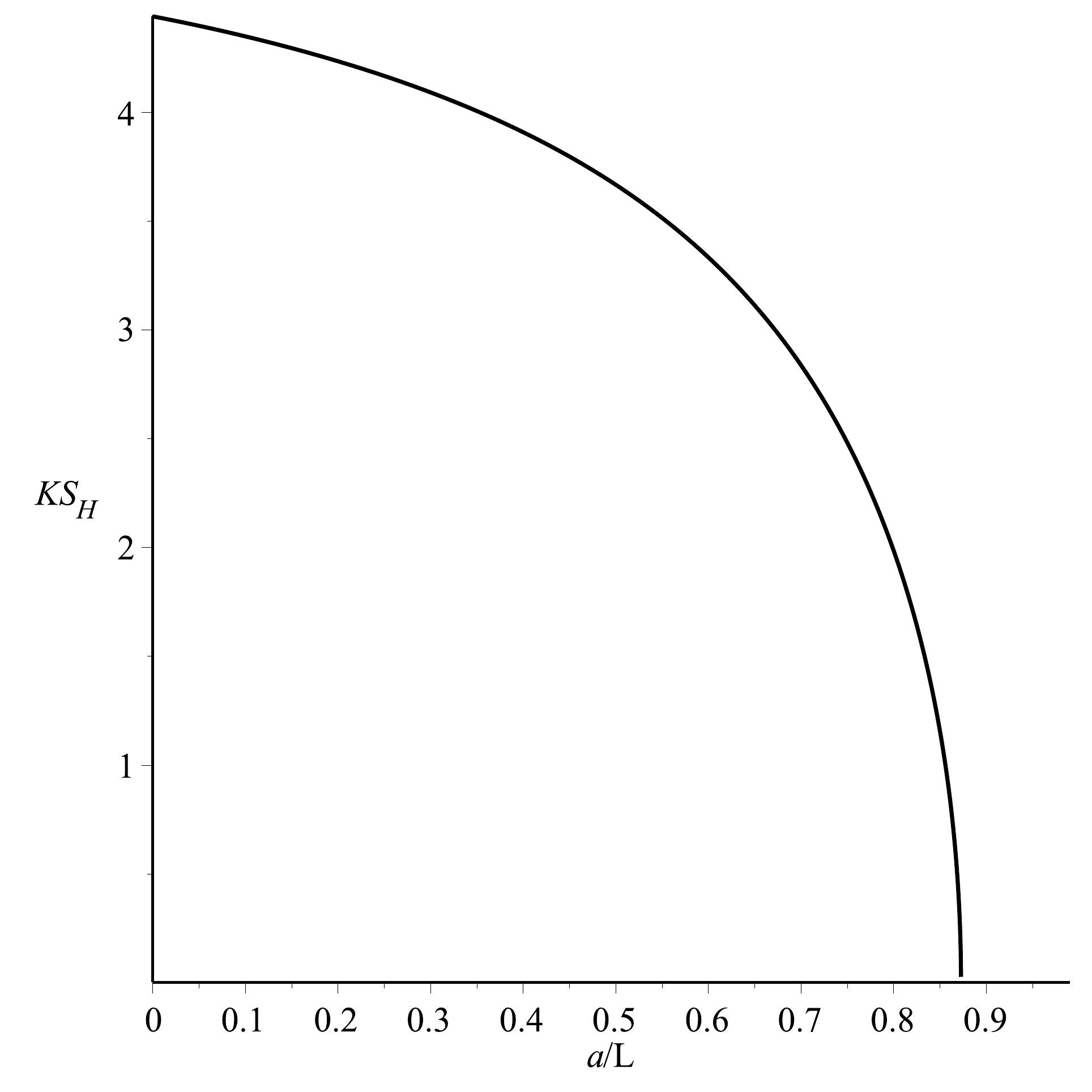}
\caption{Entropy as a function of $a/L$ for a fixed mass, $\mu = 30$, in the cisunital case ($\mathcal{A}/L < 1$). Here $K \equiv 2\ell^3_{\textsf{P}}/\pi^2$.}
\end{figure}
Notice that the entropy is maximised, for a given mass, when the black hole does not rotate; it is minimised, as usual, near ``extremality'', corresponding to a value of $a/L$ just below $0.9$. (This corresponds to $\mathcal{A}/L \approx 0.82$.)

Turning to the case where $\mathcal{A}/L > 1$: the area of the part of the event horizon outside $\theta = \theta_a$ is in like manner
\begin{eqnarray}\label{O}
\int_0^{2\pi}\int_0^{2\pi}\int_{\theta_a}^{\pi/2}{r^2_{\textsf{H}} + a^2\over |\Xi|}\,r_{\textsf{H}}\,\sin(\theta)\cos(\theta)\m{d}\theta\m{d}\phi\m{d}\psi \nonumber\\ \;=\; 2\pi^2\,{r^2_{\textsf{H}} + a^2\over |\Xi|}\,r_{\textsf{H}}\cos^2(\theta_a) \;=\; 2\pi^2\,{r^2_{\textsf{H}} + a^2\over |\Xi|}\,r_{\textsf{H}}{L^2\over a^2},
\end{eqnarray}
where we have used (\ref{M}).

Now inside $\theta = \theta_a$, $\Delta_{\theta}$ is negative; so when we cross over into that region of the event horizon, the signature of $h_{\textsf{H}}$  becomes ($-$, $-$, $-$, +), with the plus sign corresponding to $\psi$. This is of course a Lorentzian geometry (with a ``mostly minus'' signature), so $\psi$ is interpreted as time; $\theta$, $t$, and $\phi$ are spacelike. This is quite natural, since, as we saw, all three of $\theta$, $t$, and $\phi$ were spacelike on the event horizon even when $\theta > \theta_a$. All that has changed, for them, is the signature convention. In short, the only true signature ``flip'' in this case is that for $\psi$, which now represents time.

This time coordinate is clearly periodic. There is of course a large literature, and a continuing discussion, on the question as to whether closed timelike worldlines can ever be physical: see \cite{kn:kip,kn:matt,kn:caldaklemm,kn:calda,kn:martin,kn:aref,kn:rov} for a sample, presenting a wide variety of points of view; see also \cite{kn:gib1,kn:gib2} for discussions of the related issue of signature change. Note also that AdS itself, as originally defined, has a periodic time coordinate, with length $2\pi L$: see \cite{kn:gg} for the argument that this need not, and perhaps should not, be ``unwound''.

In the present case, holography offers a novel perspective on this question. At conformal infinity (with a natural choice of conformal gauge), the length $\tau$ of any closed timelike curve parametrised by $\psi$ satisfies $\tau \geq 2 \pi L^2/a$, which, by the inequality (\ref{H}), implies $\tau \geq \left(2\pi/\sqrt{2}\right)L \approx 4.44 L$. However, in the AdS/CFT correspondence, it is essential that $L$ should be very large relative to the other fundamental length scales (the five-dimensional Planck length, and also the string length scale): see \cite{kn:99}. Thus, these closed timelike worldlines are very long, suggesting from still another point of view that these objects may not be completely unacceptable. For example, if the strongly coupled system at infinity resembles the quark-gluon plasma \cite{kn:casa,kn:nat,kn:bag}, then (see again \cite{kn:99}) $4.44 L$ is far longer than the entire lifetime of the plasma; so the fact that these worldlines are closed becomes irrelevant.

Resuming our computation of the (three-dimensional) area of the event horizon: if we fix ``time'' at $\psi = $ constant, the rank of $h_{\textsf{H}}$ drops to 2, so the event horizon in this region has in effect collapsed from three to two dimensions (corresponding to the vectors dual to the one-forms $\m{d}\theta$ and $a\,\m{d}t - {r^2\,+\,a^2 \over \Xi}\,\m{d}\phi$). Consequently this part of the event horizon does not contribute to the three-dimensional area\footnote{The author is grateful to Prof. Ong Yen Chin for pointing this out.}, and so the only contribution to the entropy is given by (\ref{O}).

Thus, the horizon entropy in this case is
\begin{equation}\label{P}
S_{\textsf{H}}(\mathcal{A}/L > 1) \;=\; {\pi^2\over 2 \ell_{\textsf{P}}^3}\,{r^2_{\textsf{H}} + a^2\over |\Xi|}\,r_{\textsf{H}}{L^2\over a^2}.
\end{equation}

Again, the entropy can be regarded as a function of $\mu$ and $a/L$; but in this case, it is a monotonically \emph{increasing} function of $a/L$ when $\mu$ is fixed: see Figure 3, where again $\mu = 30$.
\begin{figure}[!h]
\centering
\includegraphics[width=0.7\textwidth]{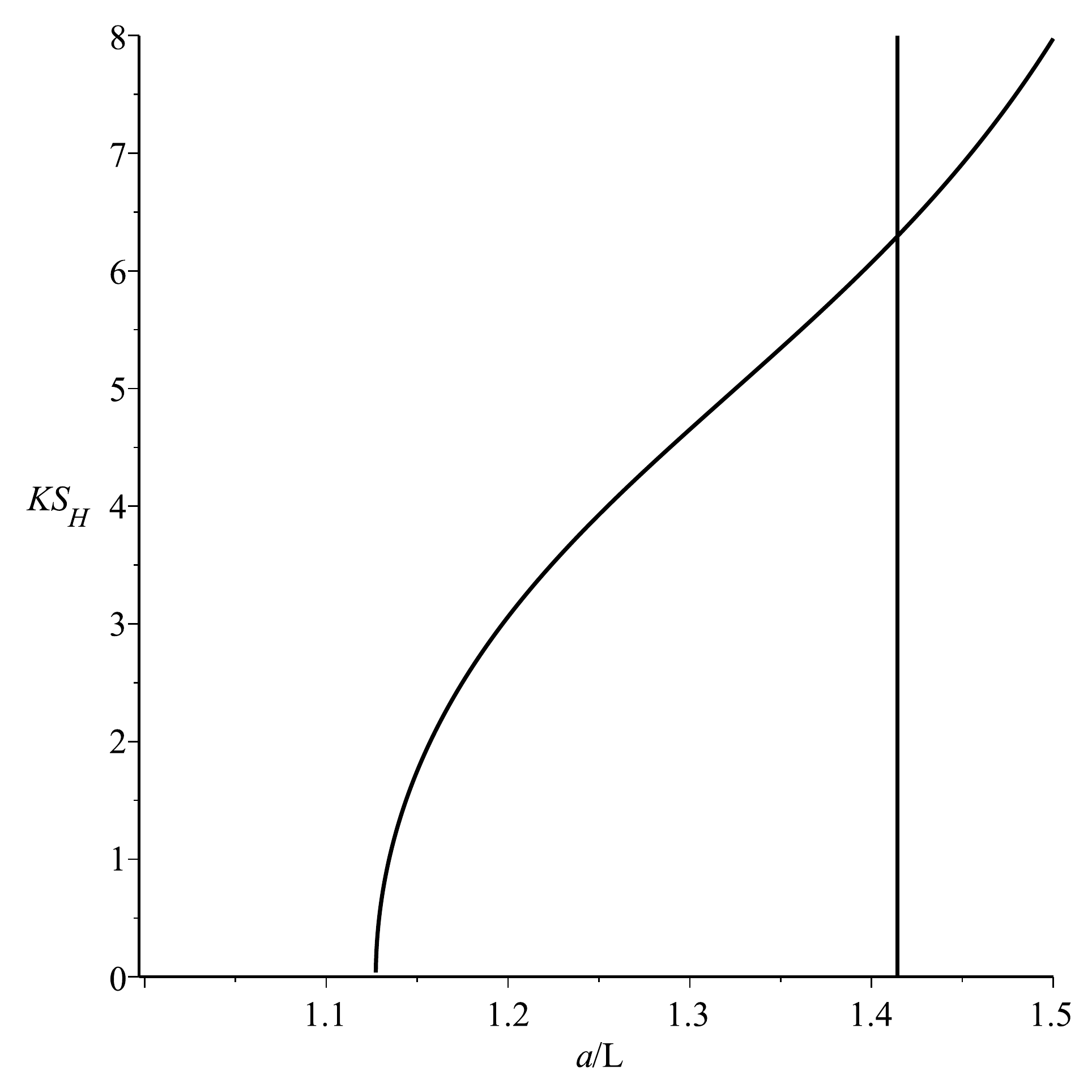}
\caption{Entropy as a function of $a/L$ for a fixed mass, $\mu = 30$, in the transunital case ($\mathcal{A}/L > 1$). Here $K \equiv 2\ell^3_{\textsf{P}}/\pi^2$.}
\end{figure}

As in the cisunital case, the entropy is very small near ``extremality'', but now that corresponds to a value of $a/L$ just \emph{above} $1.1$ (which means that $\mathcal{A}/L \approx 1.23$). It then increases, and would do so without bound if Seiberg-Witten instability did not intervene, imposing the bound indicated by the vertical line in Figure 3. Thus, in this case, entropy is maximal, for a given mass, when the black hole rotates as \emph{rapidly} as that bound permits. Notice however that the actual values attained in the permitted range are of roughly the same order as those attained in the cisunital case.

We have seen that requiring the absence of the Seiberg-Witten instability places a bound on one of the black hole parameters, $a$. Now we recall (from \cite{kn:98}) that another, completely different form of instability imposes bounds on the other parameter, $M$.

\addtocounter{section}{1}
\section* {\large{\textsf{3. Bounds from Avoidance of Superradiance}}}
Relative to a zero-angular-momentum observer at infinity, a (massless) particle, with zero angular momentum, propagating on the event horizon\footnote{Notice that we always assume that an event horizon exists. Therefore the results of this Section are based on the assumption that Censorship always holds for asymptotically AdS$_5$ black holes.} of an AdS$_5$-Kerr black hole, has an angular velocity given by
\begin{equation}\label{Q}
\omega\;=\;{a\,\left(1 + {r_{\textsf{H}}^2\over L^2}\right)\over r_{\textsf{H}}^2 + a^2}.
\end{equation}
If this angular velocity becomes too large, the system develops the well-known \emph{superradiant instability} \cite{kn:super}. Specifically, the hole can only be stable if \cite{kn:hawkreall}
\begin{equation}\label{R}
{{a\over L}\,\left(1 + {r_{\textsf{H}}^2\over L^2}\right)\over {r_{\textsf{H}}^2\over L^2} + {a^2\over L^2}}\;<\;1.
\end{equation}

Using equation (\ref{LL}) to eliminate $r_{\textsf{H}}$, we regard this expression as a function of $a/L$ and $M/L^2$. Doing so, one finds that in the cisunital case the inequality (\ref{R}) can be expressed as
\begin{equation}\label{SS}
a/L\;<\;{1\over 3}\left(1+27{M\over L^2}+3\sqrt{81{M^2\over L^4}+6{M\over L^2}}\right)^{1/3}+{1\over 3\left(1+27{M\over L^2}+3\sqrt{81{M^2\over L^4}+6{M\over L^2}}\right)^{1/3}}\,-\,{2\over 3}.
\end{equation}
The graph of this function is the lower of the two curves shown in Figure 4. The only cisunital black holes stable against superradiance are those with parameter pairs such that the corresponding point in this diagram lies \emph{below} that curve.
\begin{figure}[!h]
\centering
\includegraphics[width=0.7\textwidth]{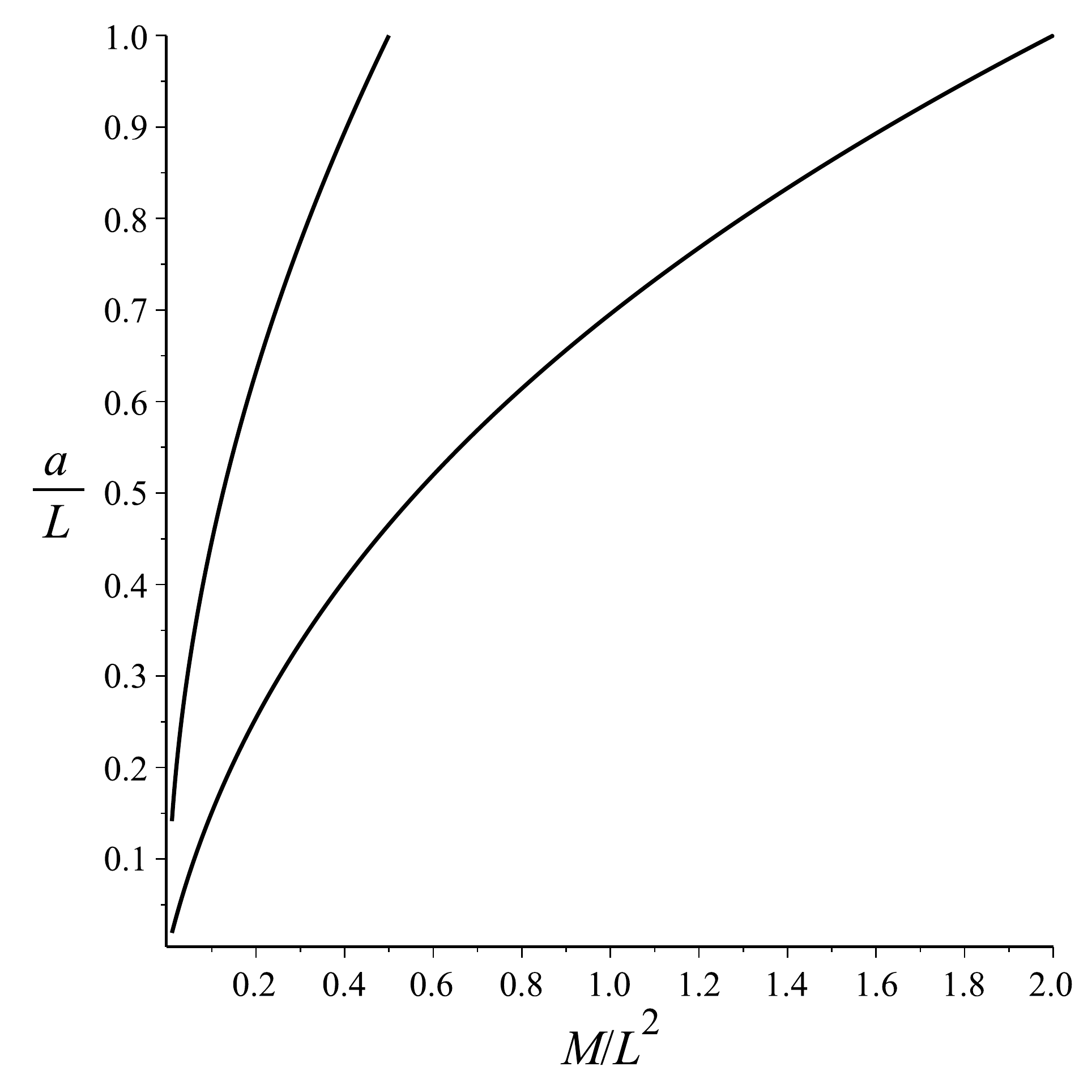}
\caption{Cisunital AdS$_5$-Kerr black holes stable against superradiance correspond to the domain below the lower curve. Those satisfying Censorship correspond to the domain below the upper curve.}
\end{figure}
The upper graph in Figure 4 corresponds to the condition for Censorship to hold. Thus, Censorship is satisfied only by those black holes with parameter pairs which lie below this upper curve. \emph{We see at once that the requirement of stability against superradiance is considerably stronger than Censorship} for these black holes.

When $\mathcal{A}/L > 1$, one finds that the inequality in (\ref{SS}) is reversed:
\begin{equation}\label{S}
a/L\;>\;{1\over 3}\left(1+27{M\over L^2}+3\sqrt{81{M^2\over L^4}+6{M\over L^2}}\right)^{1/3}+{1\over 3\left(1+27{M\over L^2}+3\sqrt{81{M^2\over L^4}+6{M\over L^2}}\right)^{1/3}}\,-\,{2\over 3}.
\end{equation}
Combining this with (\ref{H}), we see that a transunital black hole can be stable against both the Seiberg-Witten and superradiant instabilities if the parameters $a/L$ and $M/L^2$ are confined to a finite domain in the $(M/L^2, a/L)$ plane: see Figure 5.
\begin{figure}[!h]
\centering
\includegraphics[width=0.7\textwidth]{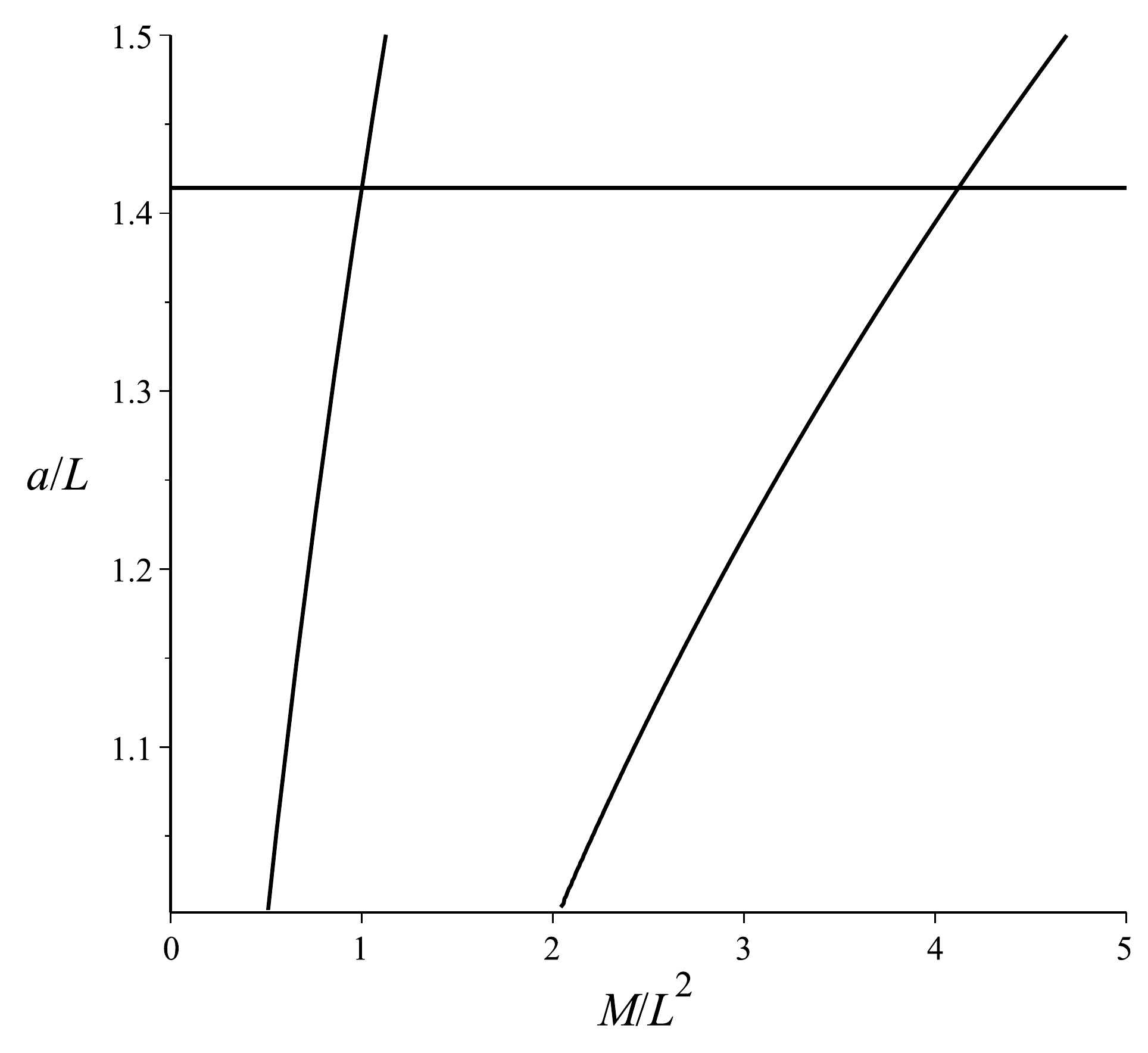}
\caption{Transunital AdS$_5$-Kerr black holes satisfying Censorship, and stable against both Seiberg-Witten and superradiant instabilities, correspond to the domain with vertices $(M/L^2 , a/L) = (0.5, 1), (1, \sqrt{2}), ((3\sqrt{2} + 4)/2, \sqrt{2}), (2, 1)$.}
\end{figure}
Notice that in this case Censorship forbids all values of $M/L^2$ below $1/2$. It constrains $a/L$ to a range between $1$ and some value strictly smaller than $\sqrt{2}$, the Seiberg-Witten bound (\ref{H}), when $1/2 < M/L^2 < 1$. For values of $M/L^2$ between 1 and 2, any value of $a/L$ between $1$ and $\sqrt{2}$ is acceptable; for higher values of $M/L^2$, up to $M/L^2 = (3\sqrt{2} + 4)/2 \approx 4.121$, an increasingly narrow range of $a/L$ values is possible; beyond that, the black hole is definitely unstable.

We see, then, that some cisunital, but also some transunital, black holes can be stable against both of the instabilities we have studied thus far; but the permitted parameter values are severely restricted.

But there is a third test. As we have seen, AdS$_5$ black holes (both cisunital and transunital) with $\mathcal{A}/L$ close to unity tend to be flattened in the equatorial plane, and this is well known to be a source of instability. We now wish to ask: can (otherwise stable) AdS$_5$ black holes satisfy the Emparan-Myers stability condition \cite{kn:empmy,kn:pau} for these flattened black holes? Our results to this point make it much easier to answer this question, because we now need only consider certain definite ranges of parameter values, corresponding to the domains shown in Figures 4 and 5.

\section* {\large{\textsf{4. Emparan-Myers Fragmentation}}}
Emparan and Myers argue that a sufficiently distorted five-dimensional black hole might \emph{fragment} into smaller black holes. The suggestion is that this will happen if it is possible to construct a pair of widely separated black holes which [a] have a total energy equal to that of the original black hole, and [b] have a total \emph{entropy} greater than that of the original black hole. The argument is that, if such a system can be constructed, then the original black hole will tend to evolve towards that state, in accordance with the Second Law of (black hole) thermodynamics. That is, the black hole will split into two pieces. This elegant argument allows us to avoid discussing the extremely complicated fission process itself.

The fragmentation may be either temporary or permanent, but in either case this is clearly a new form of black hole instability. Emparan and Myers used this idea, together with several other assumptions and approximations, to compute an upper bound on black hole specific angular momentum in the uncharged asymptotically  flat case; so of course this process is of interest to us here\footnote{We are assuming throughout this discussion that all of the black holes we consider have well-defined entropies, which means that, as in the preceding Section, we are assuming that Censorship holds for all asymptotically AdS$_5$ black holes. \emph{All} of our results are based on the idea that Censorship holds for these black holes.}.

Before we begin to discuss the details, it must be admitted that, while the thermodynamic part of this argument is clear, the additional assumptions and approximations one needs produce an explicit calculation mean that the full argument is somewhat heuristic. It is not clear, for example, that energy conservation is valid for a system, like this one, that lacks a timelike Killing field defined everywhere and at all times, including during the fragmentation process; in more familiar language, gravitational waves generated by the fragmentation will carry away some energy. We follow \cite{kn:empmy} (see its appendices) and work with the approximation that this effect can be neglected.

Again, the argument implicitly assumes that the fragmentation process is sufficiently violent that the fragments are flung to a large distance from the original location of the black hole, large enough that each fragment can eventually be treated approximately as an isolated object. (This assumption is necessary if we are to compute the entropy of the fragments.) If the original black hole spins very rapidly, as is certainly the case (for example) for transunital black holes, this is a reasonable assumption; but it is clearly not justified in the case of cisunital black holes with relatively small specific angular momenta. Even if the fragmentation is thermodynamically favoured, it can only occur if the conservation laws are respected: of course, a sufficiently slowly rotating black hole will not fragment, under any circumstances.

Since the publication of \cite{kn:empmy}, a vast amount of work has been done to put its results on a firmer basis: see for example \cite{kn:hart} for an entry point to the literature. These works have confirmed the validity of the results of \cite{kn:empmy}. Since the AdS$_5$-Kerr geometry is much more complex than that of four-dimensional, asymptotically flat black holes, even the most basic discussion of fragmentation in this case is rather intricate, as we are about to see; so we will work here at the level of rigour of \cite{kn:empmy}, leaving a more precise discussion for future work.

Let us begin with a general discussion of the complications to be expected here.

In the uncharged asymptotically  AdS$_5$ case, it is clear from Figure 2 that, if a black hole with $\mathcal{A}/L < 1$ can evolve by fragmentation, the result will be black holes with specific angular momenta as close to zero as possible; since that maximises the entropy of these objects. This was stressed by Emparan and Myers: the same phenomenon occurs in the asymptotically flat case. In short, if a cisunital black hole fragments, the resulting black holes are also (trivially) cisunital.

Because of this, if a cisunital black hole splits, its spin angular momentum is entirely converted to the orbital angular momentum of the smaller black holes. The latter therefore have non-zero linear momenta relative to the centre of mass of the system, so each of them will have relatively small masses, by the energy-momentum relation (a version of which applies also in the AdS case); that is, the mass of each fragment is substantially smaller than half of the mass of the original black hole. In some cases these masses could be quite small, but this causes no difficulties: as we saw, the fragments do not rotate and cannot violate Censorship or become unstable in any of the ways we have discussed. However, this means that the entropies of the fragments are reduced, which of course tends to suppress fragmentation itself.

On the other hand, this effect will be most pronounced when the original black hole rotates rapidly, and, for cisunital black holes, this reduces the \emph{initial} entropy (see again Figure 2)\footnote{It also affects the impact parameter of the fragmentation process, and this in turn affects the masses of the fragments because it controls the conversion of spin to orbital angular momentum.}. In short, there are two competing effects at work here, so it is difficult to say whether the total entropy of the fragments, maximal though it is \emph{given their masses}, exceeds that of the original black hole. This can be settled only by means of a detailed calculation of the masses of the fragments. However, intuition suggests that fragmentation will occur for these relatively non-exotic objects only when the initial specific angular momentum is sufficiently large. This will prove to be the case.

By contrast, if a black hole with $\mathcal{A}/L > 1$ is given the opportunity to evolve, Figure 3 shows that the result of fragmentation will be black holes with specific angular momenta as \emph{large} as Seiberg-Witten instability permits: if a transunital black hole fragments, it produces black holes which are themselves transunital. That is, part of the spin angular momentum of the original black hole is transferred to the spins of the fragments.

Since however there is a limit to the amount of angular momentum that can be taken up by the spin of the fragments, the orbital angular momentum of the fragments will still be large if the original black hole had a very large angular momentum. By the same argument as in the cisunital case, this can mean that their masses may be small. In that case $\Gamma^+_{\mu}$ (see equation (\ref{K})) will be relatively large for the fragments, and so they are close to being extremal; see the inequality (\ref{J}); the point is that the gap between the lower and upper bounds might be very narrow. However, if they are close to ``extremality'', then their entropies will be very \emph{small} relative to the entropy of the original black hole. (The entropies of the fragments are ``large'' only when compared to their masses.)

On the other hand, in contrast to the cisunital case, a large initial specific angular momentum corresponds to a large initial entropy. Paradoxically, then, this argument suggests that transunital black holes with relatively \emph{large} specific angular momenta may be stable against fragmentation, because the initial entropy is large while that of the fragments is small in this case: fragmentation is suppressed thermodynamically for these objects. This is less surprising if one recalls that it is the transunital black holes with relatively smaller specific angular momenta that have the most deformed event horizons --$\,$ recall our discussion of the event horizon circumference, above.

Now, however, there is a final complication. We recall that Seiberg-Witten instability \emph{also} limits how large the specific angular momentum of the \emph{original} black hole can be. Hence it is not obvious that the putative suppression of fragmentation by high specific angular momenta, as we have been discussing, will actually have an opportunity to occur. In short, it is not clear that otherwise stable transunital black holes can ever be immune to fragmentation. Again, this has to be settled by a detailed calculation.

In summary: we expect that cisunital AdS$_5$-Kerr black holes will fragment (into other cisunital, indeed non-rotating, black holes) unless their specific angular momenta are sufficiently \emph{small}, and that their transunital counterparts will fragment (into other transunital black holes), unless --$\,$ \emph{possibly} --$\,$ their specific angular momenta are sufficiently \emph{large}.

We now explore these expectations in detail.

\subsubsection*{{\textsf{4.1 Fragmentation when $\mathcal{A}/L < 1$ }}}
We now make explicit the Emparan-Myers fragmentation condition for cisunital AdS$_5$-Kerr black holes. We follow \cite{kn:empmy} closely; see also \cite{kn:bog} (where a different approach is used, however with broadly similar conclusions).

As explained earlier, we follow \cite{kn:empmy} in assuming that the fragments recede to large distances from the site of the original black hole\footnote{In the asymptotically flat case, Emparan and Myers assume that the fragmentation is sufficiently energetic that the fragments recede to infinity. That cannot happen in the AdS context, but in any case all we need is the assumption that the fragments recede to a distance sufficiently great that their entropies can be approximately computed as if they were isolated.}. On this scale, the fragments can be treated as point particles moving in the asymptotic AdS$_5$ spacetime geometry, with trajectories separated by an impact parameter $2R$ (at the location of the original black hole). As above, the mass and angular momentum of the original black hole are $\mathcal{M}_0$ and $\mathcal{J}_0$, while the mass and angular momentum of each fragment (separately) are $\mathcal{M}_1$ and $\mathcal{J}_1 = 0$. (Henceforth we use these subscripts, $0$ and $1$, without further explanation, to refer to quantities describing the initial and final states respectively.)

As explained earlier, in the cisunital case the entire spin angular momentum of the original black hole is transferred to the orbital angular momenta of the fragments; each has $\mathcal{J}_0/2$, so each has linear momentum $\pm \mathcal{J}_0/(2R)$. Energy conservation requires the initial total energy, $\mathcal{M}_0$, to equal the total energy of the fragments at $r = 0$. The energy-momentum relation in AdS takes the form $\mathcal{M}^2 = \alpha(r) E^2 - {p^2\over \alpha(r)}$, where $\alpha(r)$ is given by $\alpha(r) = 1 + (r^2/L^2)$. We have $\alpha(0) = 1$, and so
\begin{equation}\label{U}
\mathcal{M}_0\;=\;2\,\sqrt{{\mathcal{M}_1^2} + {\mathcal{J}_0^2\over 4R^2}};
\end{equation}
this is written more usefully as
\begin{equation}\label{V}
\mathcal{M}_1\;=\;{1\over 2}\,\sqrt{\mathcal{M}_0^2 - {\mathcal{J}_0^2\over R^2}}.
\end{equation}
Using the definition of $\mathcal{A}$, we can write this as
\begin{equation}\label{W}
\mathcal{M}_1\;=\;{1\over 2}\,\mathcal{M}_0\,\sqrt{1 - {\mathcal{A}_0^2\over R^2}}.
\end{equation}
This equation quantifies the extent to which the mass of each fragment is smaller than half the mass of the original black hole. Notice that it is not as simple as it appears, since $R$ depends on the parameters of the initial black hole, including $\mathcal{A}_0$.

We therefore need to discuss how $R$ is to be selected. Absent a detailed account of the fragmentation process, this can only be done in a somewhat heuristic way. Emparan and Myers consider a variety of candidates for $R$, guided by the intuition that it should not exceed the ``radius'' of the event horizon in the rotation plane. We have argued above that the most suitable candidate for this ``radius'' is the circumferential radius, $R_{\textsf{c}}$, defined by equation (\ref{NORMAL}), and we therefore propose $R = R_{\textsf{c}}$.

Equation (\ref{W}) shows, however, that $R$ cannot be chosen arbitrarily: clearly, \emph{any} choice of $R$ must satisfy $R > \mathcal{A}_0$ under all circumstances. We must therefore verify this for our candidate.

From equation (\ref{NORMAL}) we see that $R_{\textsf{c}}$ depends on $a_0$ and on $r_0$, the event horizon coordinate of the original black hole, which in turn depends (through equation (\ref{LL})) on $a_0$ and $M_0$. The ratio $R_{\textsf{c}}/\mathcal{A}_0$ can therefore be expressed as a function of $a_0$ and $M_0$ (or rather, of $a_0/L$ and $M_0/L^2$).

A numerical investigation of this function shows that, indeed, $R_{\textsf{c}} > \mathcal{A}_0$ for cisunital black holes (in fact, the minimum value for the ratio $R_{\textsf{c}}/\mathcal{A}_0$ is just over 3). We will see later that the corresponding statement is true of transunital black holes. Thus $R_{\textsf{c}}$ is a reasonable and mathematically well-defined choice for $R$, and we shall adhere to it henceforth. We are now in a position to compute the masses of the fragments in terms of the parameters of the original black hole.

We can now proceed to compute the ratio of the final total entropy of the putative fragments to the entropy of the original system. If $S_0$ denotes the entropy of the original black hole and $S_1$ that of each one of the fragments, this ratio is, from equation (\ref{NNN}),
\begin{equation}\label{X}
{2S_1\over S_0}\;=\;{2r_1^3\Xi_0\over r_0\left(r_0^2 + a_0^2\right)};
\end{equation}
here we have used the fact that the fragments do not rotate in this case.

Now $r_0$ is found as a function of $a_0$ and $M_0$ by solving (\ref{LL}), and similarly $r_1$ can be expressed in terms of $M_1$. We will therefore be able to express $2S_1/S_0$ in terms of $a_0$ and $M_0$ if we can do that for $M_1$.

From equation (\ref{E}) we have
\begin{equation}\label{Y}
{M_1\over M_0}\;=\;{\left(2 + \Xi_0\right)\mathcal{M}_1\over 3\,\Xi_0^2 \,\mathcal{M}_0}\;=\;{2 + \Xi_0\over 6\,\Xi_0^2}\,\sqrt{1 - {\mathcal{A}_0^2\over R_{\textsf{c}}^2}},
\end{equation}
where we have used equation (\ref{W}) and our identification of $R$ as $R_{\textsf{c}}$. As already mentioned, the latter is a known function of $a_0$ and $M_0$, so (\ref{Y}) gives us the desired expression for $M_1$. Substituting this into (\ref{X}) (after expressing $r_1$ in terms of $M_1$), we can therefore express $2S_1/S_0$ explicitly in terms of $a_0$ and $M_0$.

The expression is very complicated and not illuminating, so we set $2S_1/S_0 = 1$, so that $a_0/L$ can be regarded as a function of $M_0/L^2$. The graph of this function demarcates the parameter pairs of black holes for which thermodynamics favours fragmentation: see Figure 6, where we have superimposed the graph on Figure 4.
\begin{figure}[!h]
\centering
\includegraphics[width=0.7\textwidth]{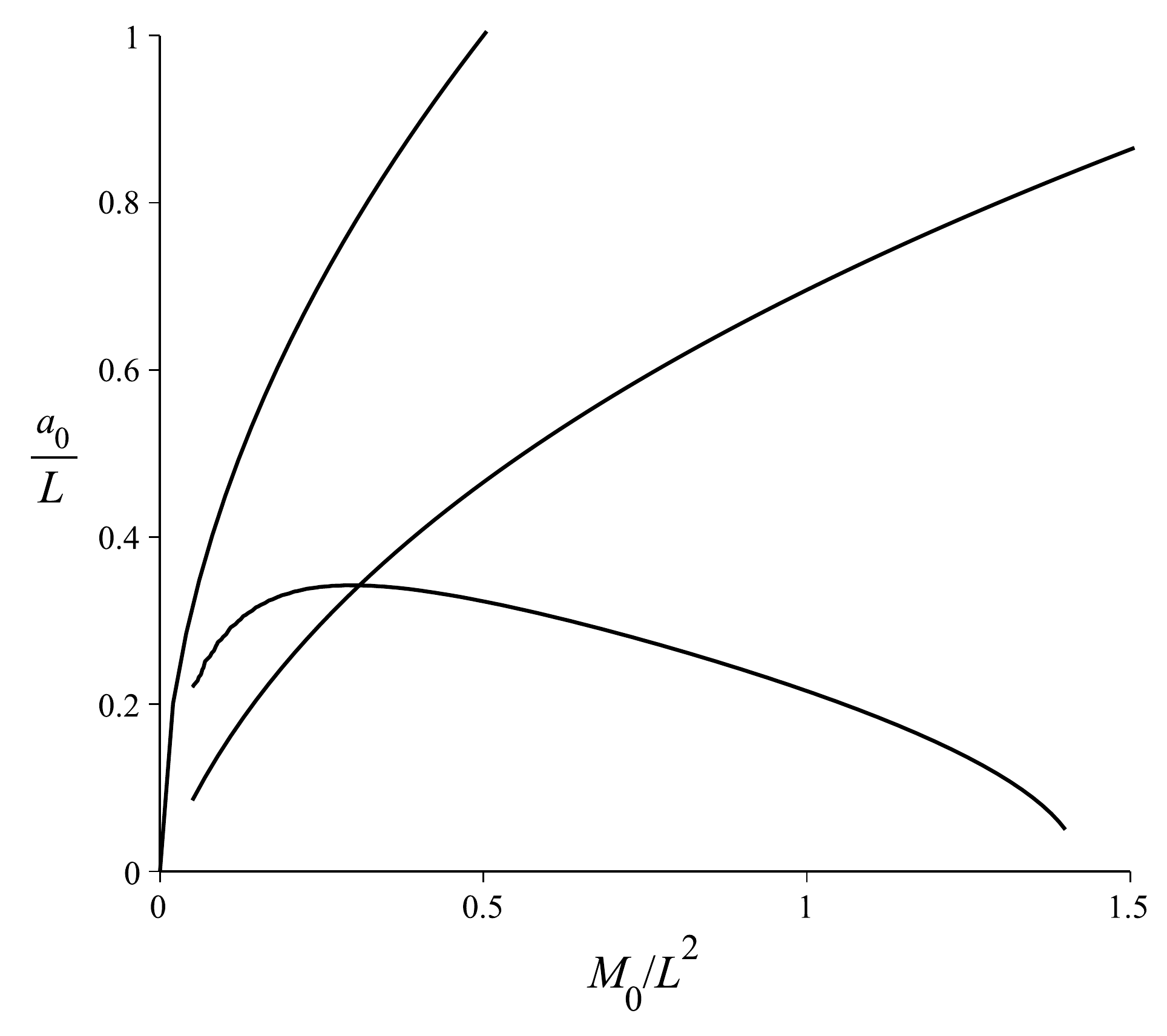}
\caption{The curve $2S_1/S_0 = 1$, for cisunital black holes (bottom), and the curve $a_0 = \sqrt{2M_0}$, the upper limit for $a_0$ given $M_0$ if Censorship is to hold (top). Between them is the curve indicating when superradiance occurs. Cisunital black holes stable against superradiance and fragmentation correspond to points lying simultaneously below the lower two curves; they automatically respect Censorship. See however the text.}
\end{figure}

The upper curve represents the upper bound on $a_0$ imposed by Censorship. The lower curve is $2S_1/S_0 = 1$. At each point, it lies (usually far) below the Censorship curve; the curves do not intersect. This means that avoidance of fragmentation is (much) more restrictive than Censorship for these black holes. Between those two curves is the curve describing when superradiance intervenes. For some values of $M_0/L^2$, the avoidance of superradiance imposes a stronger condition on $a_0/L$ than the avoidance of fragmentation, but for larger values of $M_0/L^2$ the reverse is true. Cisunital black holes respecting Censorship and suffering from neither instability have parameter pairs lying below all of these curves.

At this point, however, we must remind ourselves that this calculation is not to be trusted for very low values of $a_0/L$, since in that case there will not be sufficient initial angular momentum to ensure that the fragments recede to a distance large enough to justify our computation of their entropies. We will therefore ignore the lower section of the diagram; for example, the way the $2S_1/S_0 = 1$ curve bends down, apparently to intersect the horizontal axis, is not to be taken seriously. Conversely, the upper section of the diagram is more trustworthy. It shows that $a_0/L$ \emph{is bounded above} for fully stable cisunital black holes: we have $a_0/L \leq \approx 0.34$, which in turn means that the specific angular momentum satisfies
\begin{equation}\label{Z}
\mathcal{A}_0 \leq \approx 0.24\,L.
\end{equation}

As in our discussion of Seiberg-Witten instability, we have here a bound of a very different kind to Censorship: there is a mass-independent bound on the specific angular momentum, imposed by the requirement that the black hole should not give rise to superradiant behaviour nor break into fragments.

The lesson here is that stability against fragmentation is likely to impose conditions on black hole angular momenta which take a quite different form to the conditions imposed by Censorship, and which may well be substantially stronger than both Censorship and, in many cases, than the requirement of stability against superradiance.

We now turn to the case of transunital black holes.

\subsubsection*{{\textsf{4.2 Fragmentation when $\mathcal{A}/L > 1$ }}}
The main technical novelty here is of course the fact that it is no longer the case that all of the initial angular momentum is converted to the orbital angular momentum of the fragments. Instead, each fragment acquires an amount of angular momentum $\mathcal{J}_1$ determined by the Seiberg-Witten bound $a_1/L = \sqrt{2}$, and so equation (\ref{W}) is to be replaced by
\begin{equation}\label{ALPHA}
\mathcal{M}_1\;=\;{1\over 2}\,\mathcal{M}_0\,\sqrt{1 - {\left(\mathcal{A}_0 - \left(2\mathcal{J}_1/\mathcal{M}_0\right)\right)^2\over R^2}}.
\end{equation}
Next, we choose $R = R_{\textsf{c}}$ with the same justification as before: a numerical investigation shows that $R_{\textsf{c}} > \mathcal{A}_0$ on the domain of interest to us (from Figure 5) in this case also (the minimum of $R_{\textsf{c}}/\mathcal{A}_0$ is about $1.05$), and so the expression under the square root on the right side of (\ref{ALPHA}) must be positive, with this choice.

As before, $R_{\textsf{c}}$ can be expressed in terms of $a_0/L$ and $M_0/L^2$. Next, from equation (\ref{F}) we have $\mathcal{J}_1\;=\;\pi M_1L/\left(\sqrt{2}\ell_{\textsf{P}}^3\right)$. Combining this with equation (\ref{E}), we find
\begin{equation}\label{BETA}
{2\mathcal{J}_1\over \mathcal{M}_0}\;=\;{4\sqrt{2}L\Xi_0^2M_1\over \left(2 + \Xi_0\right)M_0}.
\end{equation}
Substituting this into (\ref{ALPHA}) and again using (\ref{E}) to express $\mathcal{M}_1/\mathcal{M}_0$ in terms of $a_0$ and $M_1/M_0$, we obtain an equation which can be solved for $M_1$ in terms of $a_0/L$ and $M_0/L^2$.

The equation for the entropy ratio in this case, replacing equation (\ref{X}), is (from equation (\ref{P}))
\begin{equation}\label{GAMMA}
{2S_1\over S_0}\;=\;{r_1\left(r_1^2 + 2L^2\right)\left({a_0^2\over L^2} - 1\right)a_0^2\over r_0\left(r_0^2 + a_0^2\right)L^2}.
\end{equation}
As in the previous case, $r_0$ is given as a function of $a_0$ and $M_0$ by solving (\ref{LL}), and similarly $r_1$ can be expressed in terms of $M_1$, which we have also just expressed as a function of $a_0$ and $M_0$. Substituting all these expressions into equation (\ref{GAMMA}), we obtain the entropy ratio as a (complicated) function of $a_0$ and $M_0$.

As foreseen, one finds that the ratio is smaller than unity, indicating that fragmentation is not favoured thermodynamically, for sufficiently large $a_0/L$ (but not too large $M_0/L^2$: note that, at the vertex ($M_0/L^2, a_0/L) = ((3\sqrt{2} + 4)/2, \sqrt{2})$ in Figure 5, it is approximately equal to $0.981$.) All values of $a_0/L$ close to unity lead, by contrast, to fragmentation.

The key point here is that values below unity are attained \emph{before} Seiberg-Witten instability for the original black hole can set in: that is, before $a_0/L$ reaches $\sqrt{2}$. This could not have been foreseen. It means that there is a (small) set of parameter values such that the corresponding transunital black hole satisfies Censorship and is stable against all of the instabilities discussed in this work.

To be more precise about this, we set $2S_1/S_0 = 1$ as before, and obtain the curve in Figure 7.
\begin{figure}[!h]
\centering
\includegraphics[width=0.7\textwidth]{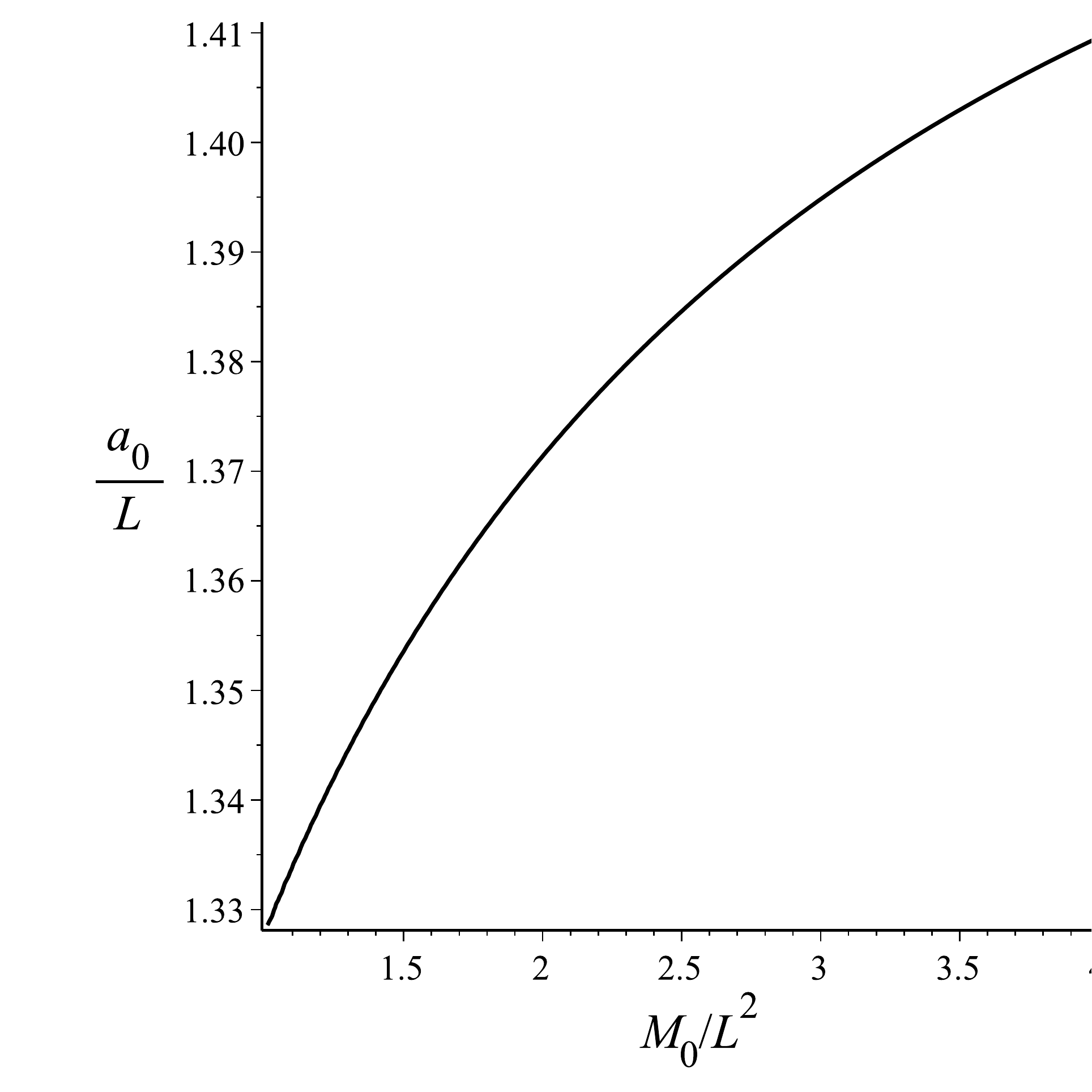}
\caption{Part of the intersection of $2S_1/S_0$, as a function of $a_0/L$ and $M_0/L^2$, with the plane $2S_1/S_0 = 1$. The region above this curve corresponds to transunital black holes stable against fragmentation.}
\end{figure}

The only transunital black holes which can resist fragmentation are those with geometric parameters ($M_0/L^2, a_0/L$) corresponding to points in this diagram \emph{above} the curve. This rules out a large part of the domain pictured in Figure 5. Indeed, combining Figures 5 and 7 we obtain Figure 8.
\begin{figure}[!h]
\centering
\includegraphics[width=0.7\textwidth]{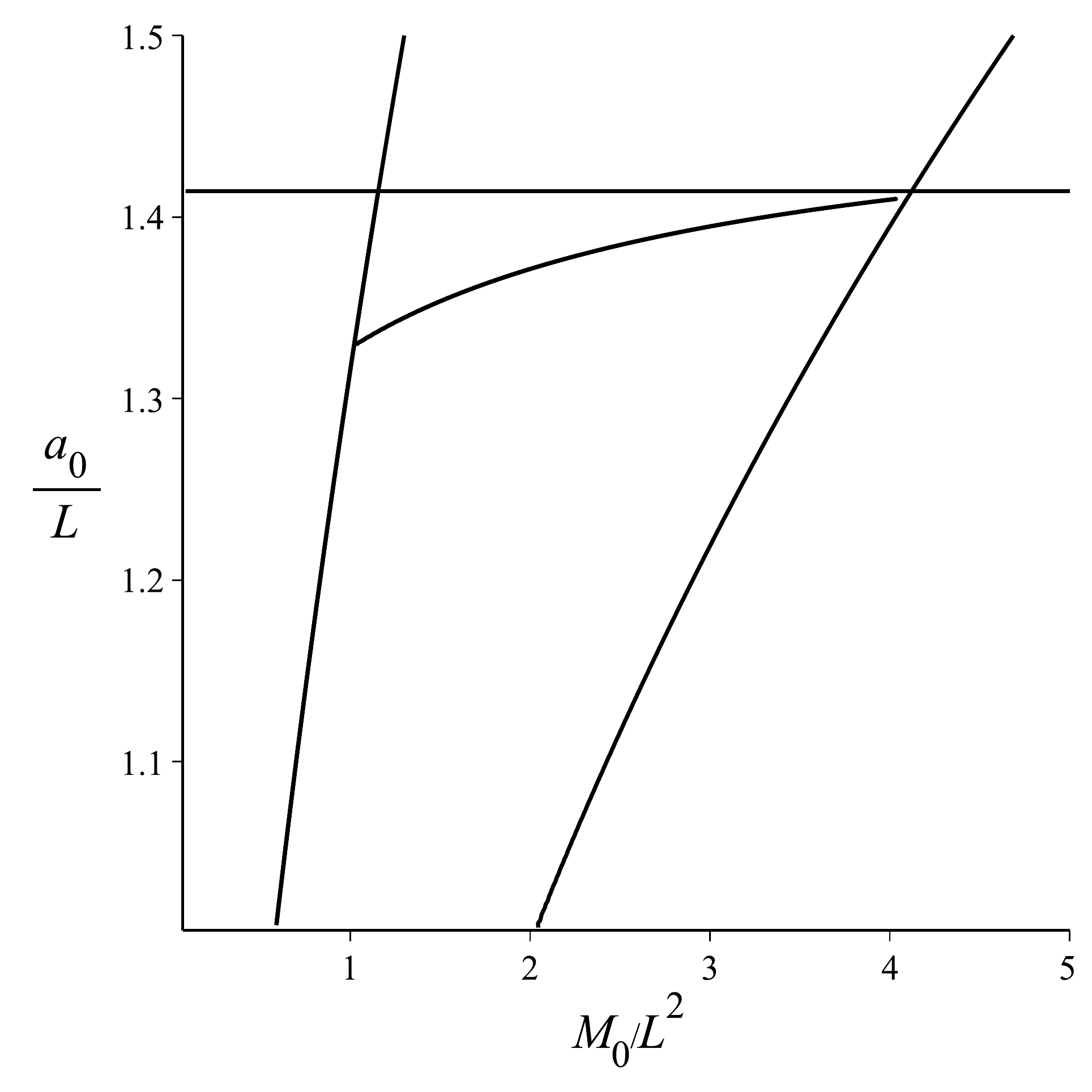}
\caption{Transunital AdS$_5$-Kerr black holes satisfying Censorship, and stable against Seiberg-Witten, superradiant, and Emparan-Myers instabilities, correspond to the small domain at the top, adjacent to the horizontal line $a/L = \sqrt{2}$, with vertices $(M/L^2 , a/L) \approx (0.88, 1.33), (1, \sqrt{2}), (4.12, \sqrt{2}), (4.08, 1.41)$.}
\end{figure}

The domain corresponding to transunital black holes which do not fragment is the upper region in Figure 8, with the indicated vertices. (The domain has four vertices, but looks triangular because two of the vertices are very close together.) Most of the region in Figure 5 with smaller values of $a_0/L$ has been eliminated, leaving a small region adjacent to the line $a_0/L = \sqrt{2}$. It is worth noting that the black holes in this category with the smallest specific angular momenta are those corresponding to the vertex $(M_0/L^2 , a_0/L) = (0.88, 1.33)$; these have $\mathcal{A}_0/L \approx 2.16$.

Bringing all of these results together: the study of black hole fragmentation indicates that the specific angular momenta of stable uncharged asymptotically  AdS$_5$ black holes are strongly restricted. Figure 6 indicates that cisunital black holes have $\approx 0.24$ as the largest possible value of $\mathcal{A}/L$, while from Figure 8 we can deduce that $\approx 2.16$ is the smallest possible value for $\mathcal{A}/L$ in the transunital case. (Of course, the largest possible value in this case is $2\,\sqrt{2} \approx 2.83$.) In this second case, the mass of the black hole is also strongly restricted\footnote{The dimensionless mass ranges from a minimum of $2$ at $(M/L^2 , a/L) = (1, \sqrt{2})$, to a maximum of $\approx 8.46$ at $(M/L^2 , a/L) \approx (4.08, 1.41)$.}. In both cases, the bounds on the specific angular momentum involve $L$, the asymptotic AdS curvature length scale, \emph{not the mass}.

\section* {\large{\textsf{5. Conclusion}}}
We have considered the proposition that asymptotically AdS$_5$ black holes of given mass cannot have arbitrarily high specific angular momenta. We have found that this is indeed the case; however, the restrictions take a complex form, and are due to Censorship only indirectly (in the sense that the analysis assumes the validity of Censorship for AdS$_5$ black holes). In particular, we have found that the requirement of stability against Emparan-Myers fragmentation imposes strong constraints: for black holes with $\mathcal{A}/L < 1$, it requires relatively very small specific angular momenta, whereas for black holes with $\mathcal{A}/L > 1$, it forces the specific angular momentum to be high, as high as \emph{another} potential instability, the Seiberg-Witten instability, permits.

Interpreting bulk physics holographically is always difficult, since it is hard to be sure to what extent the field theories on the boundary can mimic four-dimensional physics. The main lesson we would have the reader draw from this discussion is that, for systems which can be modelled in this way to some extent realistically, \emph{there are almost certainly strong bounds on the specific angular momentum}.

A key role in this work has been played by the black holes we have named ``transunital''. These may be related to the black holes with $\mathcal{A}/L \rightarrow 1$ studied in \cite{kn:klem,kn:hen,kn:supe}; for example, both types have event horizons with the same unconventional topology (a sphere with two punctures). A comparison of the two cases could be of interest.

\addtocounter{section}{1}
\section*{\large{\textsf{Acknowledgement}}}
The author is grateful to Professor Ong Yen Chin and to Dr. Soon Wanmei for useful discussions.

\end{document}